\documentclass[conference]{IEEEtran}

\usepackage{gensymb}
\usepackage{fancyhdr}
\usepackage{amsmath}
\usepackage{amssymb}
\usepackage{graphicx}
\usepackage{booktabs}
\usepackage{amsfonts}
\usepackage{multirow}
\usepackage{algorithm}
\usepackage{algorithmic}
\usepackage{mathrsfs}
\usepackage{bm}
\usepackage{exscale}
\usepackage{relsize}
\usepackage{setspace}
\usepackage{color}
\usepackage{comment}
\usepackage{ulem}
\usepackage{cite}
\usepackage{subfigure}
\usepackage{geometry}
\newtheorem{theorem}{\textbf{Theorem}}
\newtheorem{remark}{Remark}

\ifCLASSINFOpdf
\else
\fi

\begin{document}

\title{Chirp-based Hierarchical Beam Training for Extremely Large-Scale Massive MIMO}

	\author{
	Xu~Shi\textsuperscript{1},~Jintao~Wang\textsuperscript{1,2},~Zhi~Sun\textsuperscript{1},~Jian~Song\textsuperscript{1,2}\\
	\IEEEauthorblockA{
		\textsuperscript{1}Beijing National Research Center for Information Science and Technology (BNRist),\\
		Dept. of Electronic Engineering, Tsinghua University, Beijing, China\\
		\textsuperscript{2}Research Institute of Tsinghua University in Shenzhen, Shenzhen, China\\
		\{shi-x19@mails., wangjintao@, zhisun@, jsong@\}tsinghua.edu.cn}}

\maketitle

\begin{abstract}
XL-MIMO promises to provide ultrahigh data rates in Terahertz (THz) spectrum. However, the spherical-wavefront wireless transmission caused by large aperture array presents huge challenges for channel state information (CSI) acquisition. Two independent parameters (physical angles and transmission distance) should be simultaneously considered in XL-MIMO beamforming, which brings severe overhead consumption and beamforming degradation. To address this problem, we exploit the near-field channel characteristic and propose one low-overhead hierarchical beam training scheme for near-field XL-MIMO system. Firstly, we project near-field channel into spatial-angular domain and slope-intercept domain to capture detailed representations. Secondly, a novel spatial-chirp beam-aided codebook and corresponding hierarchical update policy are proposed. Theoretical analyses and numerical simulations are also displayed to verify the superior performances on beamforming and training overhead.
\end{abstract}


\IEEEpeerreviewmaketitle

\section{Introduction} 

As the communication frequency-band is further extended to millimeter-wave (mmWave) and Terahertz (THz) spectrum, extremely large-scale massive MIMO (XL-MIMO) with significant number of antennas is promising to provide much stronger beamforming gain and higher spectrum efficiency \cite{XLMIMO_3}. However, caused by the large aperture arrays and corresponding high frequency band in XL-MIMO, Rayleigh distance may appear up to several hundred meters, which means the base station (BS) will serve large near-field (i.e., Fresnel region) areas inside Rayleigh distance \cite{nearfield_1}. Spherical-wavefront assumption instead of conventional planar wavefront should be reconsidered in near-field scenario. The corresponding channel is characterized by two independent parameters, i.e., the angle-of-departure/arrival (AoD/AoA) and transmission distance \cite{nearfield_distance_2}.Therefore the transceiver, beamforming codebooks and CSI acquisition should be all redesigned for correct pair-matching of wireless channel. Moreover, the extremely large antennas and distance-sensitive channel steering response will herein bring about severe training overhead. Thus how to design the near-field beamforming codebook for XL-MIMO with low training overhead is urgent and significant. 

Conventional far-field angular-domain beam training has been widely developed in both academic research and standardization progress. The reference \cite{FFhierarchical_1_JOINT} utilized hierarchical weighted summation of sub-arrays and proposed a joint sub-array and de-activation (JOINT) hierarchical codebook. Enhanced JOINT (EJOINT) method was further proposed in \cite{FFhierarchical_2_EJOINT} to avoid antenna de-activation. Furthermore, Riemannian optimization-based method \cite{FFhierarchical_3_manifold} and successive closed-form (SCF) algorithm \cite{FFhierarchical_4_SCF} were also adopted for efficient angular coverage and partition. However, notice that the near-field XL-MIMO beam training is completely different and seems more challenging. Given that additional distance dimension should be searched, more complicated codebook and corresponding hierarchical update policy should be further developed. And the overhead consumption and computational complexity also turn extremely demanding.

However, up to now, there exists limited research for near-field hierarchical beam searching. To our best knowledge, the only existing related work for XL-MIMO is \cite{weixiuhong}, where the codewords are determined by a pair of uniformly sampled points in realistic space coordinate system. Hierarchical layers are controlled via different lengths of sampling spacing. This method is intuitive but not optimal unfortunately. The inter-beam interference and relevance are not thoroughly considered or analyzed in \cite{weixiuhong}. Consequently, nearby regions of BS may suffer from insufficient resolution due to its distance sensitivity while distance-insensitive far-field regions will be deployed with redundant codewords. The unfair distance-based codebook causes severe searching precision degradation and unnecessary training overhead. Furthermore, the codebook size will sharply boost as transmission distance raises, which is unacceptable for realistic communication. 

In this paper, we provide \emph{joint spatial-angular} and \emph{slope-intercept} representations for near-field spatial non-stationary channel. Inspired by Joint Time-Frequency Analysis (JTFA) and linear frequency modulation signal, we project the near-field beam steering vector into $2$-dim spatial-angular plane and obtain its spatial non-stationary characteristic. Just like that far-field beam can be mapped into one point in $1$-dim beamspace, we can project each near-field steering vector into one point at $2$-dim slope-intercept domain. Thus overall uniformly quantized points in slope-intercept domain are collected into one group as the elementary codebook for XL-MIMO beam training. Besides, motivated by the characteristic of chirp signal, we propose one chirp-based hierarchical beam training scheme for near-field XL-MIMO and give the detailed hierarchical update policy. The spatial-chirp beam and its beam pattern in slope-intercept domain are fully exploited in the novel training scheme. Fortunately, the novel training method can approach high beamforming gain and sum-rate with quite low training overhead.

\section{System Model}

\begin{figure}[!t]
\centering
\includegraphics[width=0.6\linewidth]{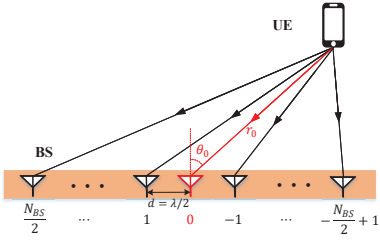}
\caption{Block diagram of near-field XL-MIMO system model.}
\label{system_model}
\end{figure}

We consider a narrow-band downlink XL-MIMO system. Base station (BS) is configured with $N_\text{BS}$-element uniform linear array (ULA) with antenna indices as $n\in \{-\frac{N_\text{BS}}{2}+1,\dots,\frac{N_\text{BS}}{2}\}$ . Without loss of generality, we only consider one single-antenna user equipment (UE) as shown in Fig. \ref{system_model}. The distance and corresponding AoD between UE and BS array center (i.e., antenna element $n=0$) are marked as $r_0$ and $\theta_0$, respectively. The carrier frequency is set as $f_c$. Denote $d=\lambda/2$ as the fixed antenna spacing and $\lambda=c/f_c$ is the wavelength of electromagnetic waves. The overall ULA size is marked as $D=d(N_\text{BS}-1)\approx dN_\text{BS}$. Therefore, the received noisy signal at UE can be formulated as 
\begin{equation}
y_t = \mathbf{h}^T\mathbf{f}_t s_t+n_t,
\end{equation}
where $\mathbf{f}_t\in \mathbb{C}^{N_\text{BS}\times 1}$ represents phase shifter (PS)-aided beamforming vector in the $t$-th timeslot and $n_t\in \mathcal{CN}(0,\sigma_N^2)$ is additive white Gaussian noise (AWGN) with power $\sigma_N^2$. $\mathbf{h}$ denotes the Terahertz wireless channel and is written as 
\begin{equation}
\mathbf{h} = \beta_\text{LoS}\mathbf{a}_\text{LoS}+ \sum_{l=1}^{N_\text{NLoS}}\beta_l \mathbf{a}_l,
\end{equation}
$\beta$ here is complex path loss and $\mathbf{a}$ is the corresponding array steering vector with each entry $a_n=e^{-j2\pi r_n/\lambda}$. For LoS path, $r_n$ denotes the distance between the $n$-th BS antenna and UE, while for NLoS path, $r_n$ represents the distance between scatterer and $n$-th BS antenna.

The main difference between XL-MIMO channel and conventional MIMO lies in the pattern of beam steering vector $\mathbf{a}_\text{LoS}$ as follows. Since the carrier frequency $f_c$ and antenna number turn extremely large, the Rayleigh length $r_\text{R}=\frac{2D^2}{\lambda}$ further increases and even gets larger than realistic supporting communication distance $r_0$, which means that the far-field plane-wave assumption couldn't hold anymore. The transmission distance for $n$-th antenna element should be exactly calculated via cosine rule as:
\begin{equation}
\arraycolsep=1.0pt\def\arraystretch{1.2}
\begin{array}{lll}
r_n &=& \sqrt{r_0^2+(nd)^2+2 r_0 nd\theta_0}\\
&\overset{(a)}{\approx}& r_0+\theta_0 \cdot nd+\frac{1-\theta_0^2}{2r_0}\cdot (nd)^2
\end{array},
\end{equation} where $(a)$ is approximated via Taylor Expansion, which has been widely adopted in previous near-field spherical-wave propagation model \cite{nearfield_1}. Notice that the first term is common to all antenna elements and is neglected here, the second term corresponds to conventional plane-wave array and the third term here is an additional component in spherical-wave XL-MIMO. Therefore the XL-MIMO array steering response $a_n=e^{-j2\pi r_n/\lambda}$ is approximate to
\begin{equation}
\arraycolsep=1.0pt\def\arraystretch{1.2}
\begin{array}{lll}
\hat{a}^\text{n-f}_n &=& \text{exp}\left\{-j\pi\left (\theta_0 n+\frac{\lambda (1-\theta_0^2)}{4r_0}n^2\right)\right\}.\\
&=& \text{exp}\left\{-j\pi\left (\theta_0 +\frac{\lambda (1-\theta_0^2)}{4r_0}n\right)\cdot n\right\}
\end{array}
\label{app_near_field_beam}
\end{equation}

The objective of beam training is to select the optimal codeword $\mathbf{f}_\text{opt}$ from finite codebook $\mathcal{F}$ to maximize the system spectral efficiency or beamforming gain. And the problem can be formulated as follows:
\begin{equation}
\arraycolsep=1.0pt\def\arraystretch{1.2}
\begin{array}{rlc}
\displaystyle \max_\mathbf{f} && |\mathbf{h}^T\mathbf{f}|^2\\
\text{s.t.} &&\ \  \mathbf{f}\in \mathcal{F}\ \  \text{and}\ \  |\mathbf{f}_n|=1,\forall n
\end{array}.
\end{equation}
Generally, the LoS path gain $\beta_\text{LoS}$ is much larger than NLoS' gain $\beta_l$ especially when carrier frequency is high enough such as mmWave and THz scenario, and thus in beam training we mainly focus on the strongest LoS beam searching.

\section{Elementary Codebook Design and $k-l$ Domain Representation}

\begin{figure}[!t]
\centering
\includegraphics[width=0.5\linewidth]{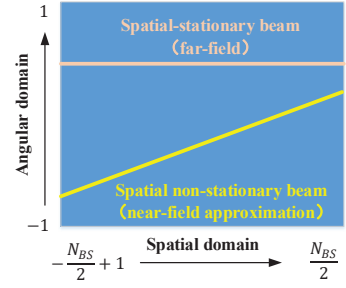}
\caption{Comparison of joint spatial-angular analysis for spatial stationary/ non-stationary beam vectors.}
\label{non_stationary_beam}
\end{figure}

Inspired by linear frequency modulation (LFM) signal $e^{-j\pi (f_0+kt)t}$ (also named as chirp signal) in continuous wave radar, we can easily observe that the approximated near-field  beam (\ref{app_near_field_beam}) has the same structure. The instantaneous AoA at each antenna element $\theta_n = \theta_0 +\frac{\lambda (1-\theta_0^2)}{4r_0}n$  increases linearly with slope $k$ and intercept $b$ as:
\begin{equation}
k=\frac{\lambda (1-\theta_0^2)}{4r_0}, \ \ \ b=\theta_0.
\label{k_b_formula}
\end{equation}
In XL-MIMO model, we have to simultaneously estimate both the slope $k$ and intercept $b$ to determine the optimal spatial-chirp beam (\ref{app_near_field_beam}), which causes more pilot consumption and tremendous challenge in hierarchical codebook design. Different from previous studies that mainly focus on direction-distance-based codebook design, we herein decouple the two parameters and consider the direct quantization of $k$ and $b$. Notice that the group $(k,b)$ is equivalent to $(\theta_0,r_0)$ due to its injective property, but $(k,b)$ is more general and low-complexity because of the decoupled relationship in chirp signal.

\begin{figure}[!t]
\includegraphics[width=1\linewidth]{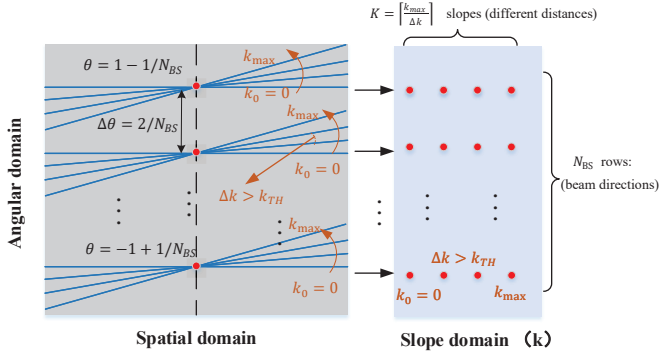}
\caption{Elementary codebook for near-field XL-MIMO in spatial-angular domain and slope-intercept ($k$-$b$) domain representation.}
\label{k_b_representation}
\end{figure}

Then a trivial and elementary codebook can be generated by uniform quantization of $k$ and $b$ as shown in Fig. \ref{k_b_representation}. First the intercept interval $[-1,1]$ are uniformly quantized to $N_\text{BS}$ groups. Inside each quantized intercept $b_q$, several quantized slopes are independently modulated to form different chirp signals. The slope interval is marked as $[k_\text{min},k_\text{max}]$ where $k_\text{min}=0$ corresponds to maximum transmission distance $r\rightarrow +\infty$ and $k_\text{max}=\frac{\lambda}{4r_\text{min}}$ corresponds to the minimum BS serve distance $r_\text{min}$. The slope quantization spacing is defined as $\Delta k<k_\text{TH}$.

Next, we transform the spatial-angular domain into slope-intercept ($k$-$b $) plane as shown in Fig. \ref{k_b_representation}. Each spatial-chirp beam (line in Fig. \ref{k_b_representation}) is projected to one point with coordinate $(k,b)$ and the whole codebook is just the uniform quantization of the 2-dim plane $[-1,1]\times [k_\text{min},k_\text{max}]$, defined as $\mathcal{F}_\text{ele}=\{\mathbf{w}_{k,b}\}$. A straightforward method for CSI acquirement in XL-MIMO is the exhaustive beam training, but we can obviously observe that the codebook size is quite huge and unacceptable for realistic traversal searching. For example, when $N_\text{BS}=1024$, $f_c= 100 \text{GHz}$ and $r_\text{min}=10\text{m}$, the corresponding codeword number is $1024\times 16=16,384$. Therefore the hierarchical spatial-chirp beam training and corresponding codebook design scheme are necessary for low pilot consumption. For the slope-intercept ($k$-$b$) domain, the following property can be easily yielded, which is helpful for the following derivations:
\begin{theorem}
Given any steering vector $\mathbf{v}\in \mathbb{C}^{N_\text{BS}\times 1}$ with unit-modulus entries ($|v_n|=1$), for $k_0$-th column's overall $N_\text{BS}$ normalized codewords ($\mathbf{w}_{k,b}$ with all uniformly quantized $b_q\in [-1,1]$ and fixed $k=k_0$, $\|\mathbf{w}_{k,b}\|^2_F=1$), the summation of coherence square is constant and equal to
\begin{equation}
\sum_{b_q}|\mathbf{w}_{k_0,b_q}^H\mathbf{v}|^2=N_\text{BS}, \ \forall k_0\in [k_\text{min},k_\text{max}]
\end{equation}
\label{Theorem1}
\end{theorem}
\begin{IEEEproof}
For any column of $k-b$ domain with $k=k_0$, the normalized codewords insides are formulated as 
\begin{equation}
    \mathbf{w}_{k_0,b_q}=\left[\frac{1}{\sqrt{N_\text{BS}}} \cdot e^{-j\pi (k_0 n^2 + b_q n)}\right],
\end{equation}
and the coherence square summation is
\begin{equation}
	\arraycolsep=1.0pt\def\arraystretch{1.2}
	\begin{array}{lll}
    \displaystyle \sum_{b_q}|\mathbf{w}_{k_0,b_q}^H\mathbf{v}|^2 & = & \displaystyle \sum_{b_q}\left|\sum_n \left(\frac{1}{\sqrt{N_\text{BS}}}e^{-j\pi k_0 n^2 } e^{-j\pi b_q n} v_n\right) \right|^2\\
    &=& \displaystyle \sum_{b_q}\left|\frac{1}{\sqrt{N_\text{BS}}} \sum_n v_n^\prime e^{-j\pi b_q n}  \right|^2
    \end{array}
    \label{parseval_equ}
\end{equation}
where $\mathbf{v}^\prime=[v_n e^{-j\pi k_0 n^2 }]$ is auxiliary vector. Notice the final result of (\ref{parseval_equ}) is just the overall power of $\mathbf{v}^\prime$ in frequency (angular) domain. According to Parseval's theorem \cite{WC_book}, we can get that the power in frequency (angular) domain is equal to the power in time (spatial) domain. Since $\mathbf{v}^\prime$ is still with all entries unit-modulus ($|v^\prime_n|=1$), the time (spatial) domain power is $\mathbf{v}^{\prime\ H}\mathbf{v}^\prime=N_\text{BS}$ and thus we finish the proof of Theorem \ref{Theorem1}.
\end{IEEEproof}

\begin{remark}
From Theorem \ref{Theorem1} we get that: All beamforming codewords contain the same power $N_\text{BS}$ when we project it into each column of $k-l$ domain, which also means that, we cannot design such a simple codeword that only scans for a fraction in slope $k$ axis. At least, it is quite difficult to find a beamforming vector that only supports a fractional square $[k_1,k_2]\times [b_1,b_2]$ (with the rest domain's coherence all zero) inside the whole domain $[k_\text{min},k_\text{max}]\times [-1,1]$. In another word, even if such codewords are designed, the corresponding inter-codeword interference and beam training overhead may further degenerate since the rest $k$-$b$ region's power is not fully considered or exploited.
\end{remark}

\section{Chirp-based Hierarchical Beam Training}

\subsection{Spatial-Chirp Beam Pattern Analysis in $k-b$ Domain}
As well known, chirp signal is a typical broadband signal, which is widely utilized for wideband target detection. To the best of our knowledge, in previous studies for far-field beam training, chirp signal (or quasi-chirp signal) has been adopted for hierarchical codebook design to scan for a large-size angular interval, like JOINT \cite{FFhierarchical_1_JOINT}, EJOINT \cite{FFhierarchical_2_EJOINT} and beam broadening \cite{beambroadening_gff}. This is because the broadband chirp signal can be conveniently controlled via only two parameters $k$ and $b$, where $k$ determines the beam width roughly while $b$ controls the beam's central direction in far-field plane-wavefront assumption.

\begin{figure}[!t]
	\begin{center}
		\subfigure[Spatial-domain beam pattern and angular-domain beam pattern of near-field LoS channel.]{
			\includegraphics[width=0.8\linewidth]{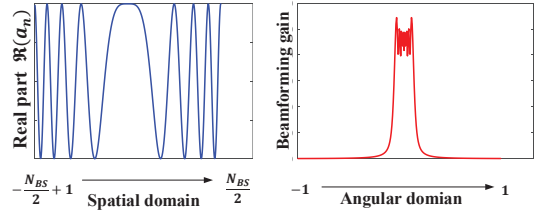}
			\label{each_spatial_angular_chirp_beam_pattern}
		}
		\subfigure[The ideal and realistic beam pattern of spatial-chirp beam in $k$-$b$ domain representation.]{
			\includegraphics[width=0.6\linewidth]{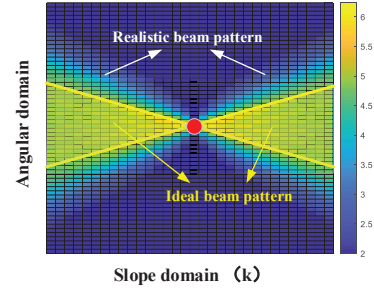}
			\label{spatial_chirp_beam_pattern}
		}
		\caption{The near-field LoS channel representation in spatial domain, angular domain and $k$-$b$ domain.}
	\end{center}
\end{figure}

Therefore, we herein give an analysis for chirp signal first, which seems significantly useful for both far-field and near-field hierarchical beam training. According to Theorem \ref{Theorem1}, chirp beamforming signal also contains the same power along all columns' coherence. But  the detailed coherence gain distribution inside $k-b$ domain is still not captured. Taking one spatial-chirp signal with $k=k_0$ and $b=b_0$ for example, the mathematical expression $\mathbf{f}_\text{eg}(k_0,b_0)$ is written in (\ref{chirp_analysis}), while the corresponding spatial-domain curve (real component) and angular-domain (k=0) curve (modulus) are shown in Fig. \ref{each_spatial_angular_chirp_beam_pattern}. 
\begin{equation}
\arraycolsep=1.0pt\def\arraystretch{1.2}
	\begin{array}{ccc}
&&\mathbf{f}_\text{eg}(k_0,b_0) =\left[e^{-j\pi\left(k_0(-\frac{N_\text{BS}}{2}+1)^2 + b_0 (-\frac{N_\text{BS}}{2}+1)\right)},\dots\right.\\
&&\left. ,1, \dots,e^{-j\pi\left(k_0(\frac{N_\text{BS}}{2})^2 + b_0 \frac{N_\text{BS}}{2}\right)}\right]^T
\label{chirp_analysis}
\end{array}
\end{equation}
Following the results in LFM signals, when with large spatial-angular product $k_0N_\text{BS}^2\gg 1$ here, the angular bandwidth of the spatial-chirp signal $\mathbf{f}_\text{eg}(k_0,b_0)$ can be easily approximated to $B_{0} = k_0N_\text{BS}$, which also means that we can coarsely scan angular interval with length $B_{0}$ via $\mathbf{f}_\text{eg}(k_0,b_0)$. Similar to the proof in Theorem \ref{Theorem1}, we can further extend it to search for one column of $k-b$ domain with $k=k_1$, and the corresponding approximated searching angular bandwidth is derived as
\begin{equation}
B_{k_1} \approx |k_0-k_1|N_\text{BS}
\label{ideal_pattern_width}
\end{equation}
When $k_1=k_0$, the scanning interval will degrade to one point, i.e, $(k_0,b_0)$ in slope-intercept plane with power $N_\text{BS}$. Notice that each beam contains a bandwidth and thus the beam point $(k_0,b_0)$  here also supports an intercept width as $\frac{2}{N_\text{BS}}$, which is consistent with unit bandwidth in orthogonal far-field discrete Fourier transform (DFT) codebooks \cite{DFTcodebook}. Therefore, we coarsely approximate the bandwidth as $B_{k_1}\approx |k_0-k_1|N_\text{BS}+\frac{2}{N_\text{BS}}$ for consistency. According to Theorem \ref{Theorem1}, the total power at each column $k=k_1$ is fixed as $N_\text{BS}$ and corresponding interval length is $B_{k_1}$. Assume the power is uniformly distributed in the interval and thus the average gain (coherence) at each inside codeword point $(b_q,k_1)$ can be yielded to $N_\text{BS}/B_{k_1}$, i.e.,
\begin{equation}
\arraycolsep=1.0pt\def\arraystretch{1.0}
	\begin{array}{lll}
\displaystyle g^\text{ideal}_{b_q,k_1}& =& \displaystyle\sqrt{ \frac{N_\text{BS}}{B_{k_1}}}\cdot \text{rect}\left(\frac{b_q-b_0}{B_{k_1}}\right)\\
&=&\left\{ \begin{array}{cl}
\displaystyle \frac{1}{\sqrt{|k_0-k_1|+2/N_\text{BS}^2}}\  &,\displaystyle \ |b_q-b_0|\leq\frac{B_{k_1}}{2}\\
0\ &, \ \text{else}
\end{array}
\right.
\end{array}
\label{ideal_pattern_gain}
\end{equation}

From above analysis we can get that, the ideal signal's searching range at $k_1$ column of $k-b$ plane is proportional to the distance ($|k_0-k_1|$) along $k$ axis, while the ideal average gain at each point inside $k=k_1$ is approximately inversely proportional to the distance $|k_0-k_1|$. Then we can depict this property into $k$-$b$ domain as shown in Fig. \ref{spatial_chirp_beam_pattern}, where the darkness represents coherence amplitude between $\mathbf{f}_\text{eg}(k_0,b_0)$ and the corresponding local point. The specific coverage pattern is extremely instructive for the hierarchical codebook design in the 2-dim spatial-chirp beam training. 

\subsection{Chirp-based Hierarchical Beam Training}

\subsubsection{Top layer hierarchical searching}
\ \ 

In the top-layer beam training, suppose the chirp-codewords are all selected at margin of $k-b$ domain $[k_\text{min},k_\text{max}]\times [-1,1]$. When the ideal beam pattern (\ref{ideal_pattern_width}) (\ref{ideal_pattern_gain}) is assumed, to fully cover the whole $k$-$b$ domain, the corresponding chirp beam distribution should be designed as shown in Layer 1 in Fig. \ref{near_field_beam_training}. Each chirp beam can support beam searching inside a delta-shaped region. The darkness also denotes beamforming gain for the corresponding channel path with coordinate at each point. Define the $i$-th codeword at column $k$ as $\mathbf{w}^{(1)}_{k,i}$, where index indicator $\ell=1$ represents the top hierarchical layer. The angular sampling spacing and codeword number at top layer should follow the next theorem:
\begin{theorem}
In top-layer beam training, angular sampling spacing is at most $B_\text{max}=\sqrt{2/N_\text{BS}}$ and thus the codeword number is at least $2\sqrt{2N_\text{BS}}$.
\label{Theorem2}
\end{theorem}
\begin{IEEEproof}
According to the near-field Fresnel region analysis \cite{minimum_distance05}, the distance lower bound of Fresnel region is $r_\text{min}=0.5\sqrt{\frac{D^3}{\lambda}}$. Therefore the minimum communication distance has $r_0>r_\text{min}$. Substituting it into (\ref{ideal_pattern_width}) we  can get that 
\begin{equation}
B_\text{max}= k_\text{max}N_\text{BS}\leq \frac{\lambda}{4r_\text{min}}N_\text{BS}=\sqrt{\frac{2}{N_\text{BS}}}.
\label{top_resolution}
\end{equation}
In the top-layer searching, the codewords are all configured at two columns ($k=k_\text{min}=0$ and $k=k_\text{max}$) with inter-column spacing $B_\text{max}$, therefore we can obtain the total codeword number at top hierarchical layer as 
\begin{equation}
N^{(1)} = 2\times \frac{1-(-1)}{B_\text{max}}=2\sqrt{2N_\text{BS}}.
\label{top_word_num}
\end{equation}
\end{IEEEproof}

From Theorem \ref{Theorem2} we can further write the coordinates of top-layer codewords as follows:
\begin{equation}
\arraycolsep=1.0pt\def\arraystretch{1.0}
\left\{	\begin{array}{llc}
\mathbf{w}^{(1)}_{k_\text{min},i}:&& \displaystyle \left(k_\text{min},iB_\text{max}\right); \\
\mathbf{w}^{(1)}_{k_\text{max},i}:&&\displaystyle  \left(k_\text{max},(i+0.5)B_\text{max}\right); 
\end{array}
\right., i=1,\dots, N^{(1)}
\label{top_layer_codeword}
\end{equation}
This conclusion is quite succinct, which shows that the top-layer codeword number and supporting region only depend on the BS antenna number $N_\text{BS}$. For example, when we set carrier frequency $f_c=30\text{GHz}$ and $N_\text{BS}=512$, the minimum communication distance $r_\text{min}$ can be easily calculated as $8\text{m}$. Thus the codeword number is obtained as $N^{(1)}=20$, which seems acceptable for realistic beam training.

\subsubsection{Hierarchical Codeword Update Policy}
\ \ 

\begin{figure*}[!t]
	\centering
	\includegraphics[width=0.7\linewidth]{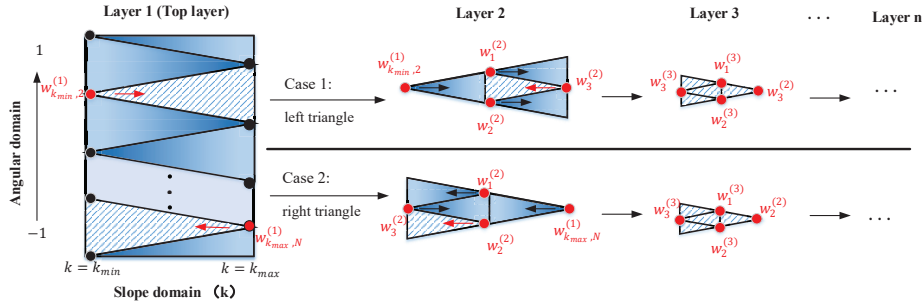}
	\caption{The hierarchical beam training and codeword update rule for near-field  ($k$-$b$ domain) XL-MIMO system.}
	\label{near_field_beam_training}
\end{figure*}

When the top-layer beam searching is addressed, we obtain the codeword with strongest beam gain which is herein marked as $\mathbf{w}^{(1)}_\text{opt}$. The corresponding $k$-$b$ domain triangle region is determined as $\mathcal{R}^{(1)}_\text{opt}$. As for each selected triangle region $\mathcal{R}^{(1)}_\text{opt}$, we divide it into four specific fractions, as shown in Layer 2 in Fig. \ref{near_field_beam_training}. We'd like to discuss the problem in two cases. For Case 1 as shown in Fig. \ref{near_field_beam_training}, the top-layer optimal region is a left triangle. Without loss of generality, we assume the optimal top-layer codeword is $\mathbf{w}_\text{opt}^{(1)}=\mathbf{w}^{(1)}_{k_\text{min},2}$. Next in Layer 2, three additional codewords are searched. The corresponding codewords' coordinates in $k$-$b$ domain can be easily obtained as middle points of triangle sides $\mathcal{R}^{(1)}_\text{opt}$, i.e.,
\begin{equation}
\arraycolsep=1.0pt\def\arraystretch{1.0}
\left\{	\begin{array}{llc}
\displaystyle \mathbf{w}^{(2)}_{1}:&&\displaystyle \left(\frac{k_\text{min}+k_\text{max}}{2},(i-0.25)B_\text{max}\right); \\
\displaystyle \mathbf{w}^{(2)}_{2}:&&\displaystyle  \left(\frac{k_\text{min}+k_\text{max}}{2},(i+0.25)B_\text{max}\right); \\
\displaystyle \mathbf{w}^{(2)}_{3}:&&\displaystyle  \left(k_\text{max},\ \ iB_\text{max}\right); 
\end{array}
\right.
\label{bottom_codeword}
\end{equation}
When in this example with $\mathbf{w}_\text{opt}^{(1)}=\mathbf{w}^{(1)}_{k_\text{min},2}$, set $i=2$ in (\ref{bottom_codeword}) and the three codewords are then obtained. Notice that the temporary codeword subset $\{\mathbf{w}^{(1)}_{k_\text{min},2},\mathbf{w}_1^{(2)},\mathbf{w}_2^{(2)},\mathbf{w}_3^{(2)}\}$ can uniformly divide region $\mathcal{R}^{(1)}_\text{opt}$ into four non-overlapping sub-triangles and each codeword dominates in one sub-region, which means by comparing the four codewords' beamforming gains, we can further reduce the potential regain by four times. Both $k$ dimension and $b$ dimension can be reduced by half. We can successively obtain $\mathcal{R}^{(3)}_\text{opt}$, $\dots$, $\mathcal{R}^{(l)}_\text{opt}$ until $l$ approaches the maximum hierarchical layer $L$. For the other case (Case 2: right triangle) in Fig. \ref{near_field_beam_training}, we emit its detailed description for brevity since its process is quite similar to Case 1. 

In retrospect, the conventional far-field beam training can be regarded as a particular case of chirp-based XL-MIMO training. As shown in Fig. \ref{near_field_beam_training}, if we only focus on the overall angular values with fixed slope $k=0$, the 2-dim hierarchical searching procedure proposed above will degrade to angular-domain 1-dim hierarchical beam training, where two additional codewords are searched in each layer. From this point of view, our proposed near-field XL-MIMO hierarchical training contains only limited pilot overhead which is realistic and quite comparable to conventional mmWave binary hierarchical beam searching.

\begin{algorithm}[htb] 
\normalem
\caption{Proposed Chirp-based hierarchical beam training for near-field XL-MIMO} 
\label{alg1} 
\begin{algorithmic}[1] 
\REQUIRE System configurations $N_\text{BS}$, $f_c$, $d$, $d_\text{min}$, required beamforming gain threshold $g_\text{th}$.

\ENSURE Beam training result $\mathbf{b}$


\STATE set $k_\text{max}=\frac{\lambda}{4r_\text{min}}$, hierarchical layer $L$, top-layer codeword number $N^{(1)}$  and angular-domain spacing $B_\text{max}=k_\text{max}N_\text{BS}$

\emph{\%\% Top-layer search}
\STATE Generate top-layer codewords via (\ref{top_layer_codeword}) and search for codeword $\mathbf{w}^{(1)}_\text{opt}$ (maximum gain) and corresponding region $\mathcal{R}^{(1)}_\text{opt}$. 

\emph{\%\% Hierarchical search}
\FOR{layer index $l=2$ to $L$}
\STATE Divide region $\mathcal{R}^{(i)}_\text{opt}$ and generate corresponding four next-layer codewords via (\ref{bottom_codeword}) 
\STATE Select the best codeword with maximum gain $\mathbf{w}^{(i)}_\text{max}$ and update the optimal region $\mathcal{R}^{i}_\text{opt}$
\ENDFOR
\STATE Final optimal supporting beam $\mathbf{b}\leftarrow \mathbf{w}^{(L)}_\text{opt}$
\end{algorithmic}
\end{algorithm}

\section{Simulation Results}

In the simulation we assume that the central frequency is $f_c=50\ \text{GHz}$ and BS antenna number is $N_\text{BS}=512$. One single-antenna user is aligned to BS via beamforming for wireless communication. The LoS path gain is calculated via the free space transmission loss, which is dependent of transmission distance and carrier frequency. The power ratio of the LoS path to the NLoS paths is denoted as $\rho=10 \text{dB}$. The minimum transmission distance can be calculated as $r_\text{min}=0.5\sqrt{\frac{D^3}{\lambda}}=12.29\text{m}$. We generate the transmission distance $r_0$ between BS array center and UE with uniform distribution $r_0\sim \mathcal{U}([13\text{m},100\text{m}])$. Similarly, the relative direction $\theta_0$ is uniformly generated insides interval $[-1,1]$. Then we can mathematically calculate the top-layer codeword number as $N^{(1)}=64$ while the overall hierarchical layer number is $L=5$. As comparison, we herein simulate several other beam training schemes. Firstly, the beamforming scheme with perfect CSI is provided as the absolute upper bound. Besides, we collect and traverse the overall chirp-based bottom-layer codewords (exhausting searching without hierarchy) for CSI acquisition as a much tighter beamforming upper bound. The far-field codebook is also compared here. Besides, we also simulate near-field distance-based searching methods \cite{weixiuhong} where transmission distance  are uniformly divided into several fractions for XL-MIMO codebook design.

\begin{figure}[!t]
	\centering
	\includegraphics[width=0.8\linewidth]{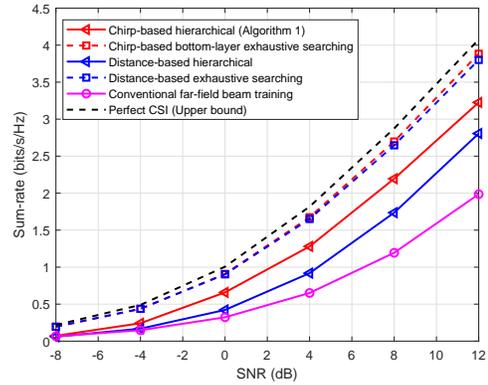}
	\caption{Comparisons of average sum-rate against SNR for several hierarchical searching schemes in near-field XL-MIMO, with $N_\text{BS}=512$ and $f_c=50\ \text{GHz}$.}
	\label{SumRate_SNR}
\end{figure}

Fig. \ref{SumRate_SNR} presents the sum-rate performances under different SNRs. We can observe that the elementary codebook (overall bottom-layer codewords) works well up to $100\%$ success rate, which demonstrates the superiority of our proposed codewords. Nevertheless, the hierarchical searching couldn't obtain absolutely $100\%$ success rate since there still exist overlapping and interference among hierarchical codewords. In fact, the ideal beam pattern is non-existent in practice and our proposed hierarchical codebook is just certain approximations of it. However, there is no doubt that the performance can be further improved via optimization, and the average sum-rate is supportive and outperforms conventional far-field training by almost $1.5\ \text{bits/s/Hz}$.

 \begin{figure}[!t]
 	\centering
 	\includegraphics[width=0.8\linewidth]{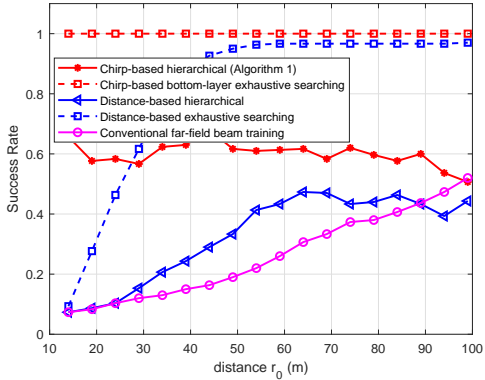}
 	\caption{Comparisons of average sum-rate against transmission distance for several hierarchical searching schemes in near-field XL-MIMO, with $\text{SNR}=10\ \text{dB}$, $N_\text{BS}=512$ and $f_c=50\ \text{GHz}$.}
 	\label{SumRate_distance}
 \end{figure}

 The impacts of transmission distance are shown in Fig. \ref{SumRate_distance}, where we fix SNR as $10\ \text{dB}$. 
 The training success rate here is defined as follows. If the accurate LoS path information is captured, or
 the beamforming gain via hierarchical searching can approach $90\%$ under perfect CSI, we regard the hierarchical scheme works successfully.
 The proposed method can support both far-filed and near-field scenarios with high sum-rate and beamforming gain, while the conventional DFT codebook fails when $r_0<100\text{m}$. In near-field scenario such as $r_0=30\text{m}$, the DFT codebook can only approach sum-rate  $1.5 \text{bits/s/Hz}$ but the proposed scheme outperforms it by almost two times. Besides, the fluctuation of chirp-based training with distance $r_0$ (red curves) is normal since the transmission distance is non-linear quantized in the near-field hierarchical codebook, while the noise and residual overlapping among codewords also affect the success rate and curve fluctuation. 
 
\begin{figure}[!t]
	\centering
	\includegraphics[width=0.8\linewidth]{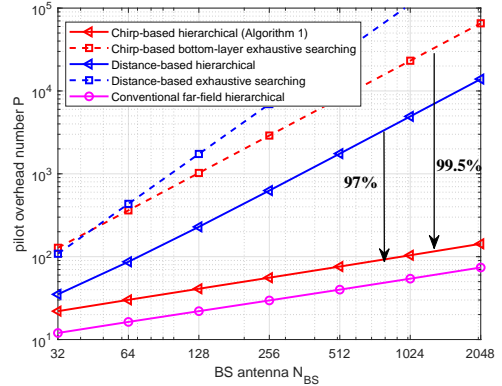}
	\caption{Overall overhead consumption against BS antenna number.}
	\label{overhead_BSantenna}
\end{figure}

We elaborate the pilot overhead consumption under different BS antenna configurations in Fig. \ref{overhead_BSantenna}. Compared with the bottom-layer overall codewords exhaustive searching, the overhead consumption can be reduced by over $99.5\%$, and over $97\%$ overhead can be saved compared with distance-based hierarchical searching method \cite{weixiuhong}. For example, when the antenna $N_\text{BS}=512$, the exhaustive searching needs overhead of about $8192$, while the chirp-based hierarchical scheme only needs $2^{2-L}N_\text{BS}+3(L-1)=76$ overhead. Therefore, we can conclude that compared with conventional DFT codebook and corresponding hierarchical schemes, our proposed near-field hierarchical searching can approach far superior and robust performance with quite comparable overhead consumption.

\section{Conclusion}

In this paper, we propose one low-overhead hierarchical beam training scheme for near-field XL-MIMO system to rapidly capture CSI with low training overhead consumption. Firstly, spatial-angular domain representation and slope-intercept domain representation are proposed to describe near-field channel. Then inspired by the LFM Radar waveform, a novel spatial-chirp beam-aided codebook and corresponding hierarchical update policy are proposed. Both training accuracy and overhead consumption can be enhanced thankfully. The near-filed hierarchical codebook enhancement will be further studied in the future.

\section*{Acknowledgment}
This work was supported in part by Tsinghua University-China Mobile Research Institute Joint Innovation Center.


%

%
%
%
%
%

\ifCLASSOPTIONcaptionsoff
  \newpage
\fi

\bibliographystyle{ieeetr}
\normalem
\bibliography{bare_jrnl.bib}

\vspace{12pt}

\end{document}